\documentclass[prl,twocolumn,showpacs,hyperref,showpacs]{revtex4-1}
\usepackage[left=1.6cm,right=1.6cm,bottom=1.6cm,top=1.6cm,ignoreall]{geometry}
\usepackage[english]{babel}
\usepackage[utf8]{inputenc}
\usepackage{epsfig}
\usepackage{color}
\usepackage{amsmath}
\usepackage{amsfonts}
\usepackage{amssymb}
\usepackage{graphicx}
\usepackage[hyperindex=true]{hyperref}
\hypersetup{linktocpage,colorlinks=true,citecolor=blue,linkcolor=blue}
\usepackage{esvect}
\usepackage[normalem]{ulem}

\makeatletter
\def\overvect@#1#2#3{\vbox{\ialign{##\crcr%
 \noalign{\kern-.7pt\nointerlineskip}#1#2\crcr%
 \noalign{\kern-0.5pt\nointerlineskip}$\m@th\hfil#2#3\hfil$\crcr}}}
\makeatother
\DeclareRobustCommand{\cev}[1]{\reflectbox{\ensuremath{\vv{\reflectbox{\ensuremath{#1}}}}}}
\DeclareRobustCommand{\Fibo}{\vv{F_N}}
\DeclareRobustCommand{\Obif}{\cev{F_N}}
\DeclareRobustCommand{\thetaa}{\vv{\theta}}

\begin{document}

\title{Measuring topological invariants from generalized edge states\\ in polaritonic quasicrystals}

\author{Florent Baboux$^{*1,2}$, Eli Levy$^{*3,4}$, Aristide Lema\^{\i}tre$^1$, Carmen G\'omez$^1$, Elisabeth Galopin$^1$, Luc Le Gratiet$^1$, Isabelle Sagnes$^1$, Alberto Amo$^1$, Jacqueline Bloch$^{1}$, Eric Akkermans$^3$}

\affiliation{$^*$These authors contributed equally to this work.}
\affiliation{$^1$Centre de Nanosciences et de Nanotechnologies, CNRS, Univ. Paris-Sud, Universit\'e Paris-Saclay, C2N Marcoussis, 91460 Marcoussis, France}
\affiliation{$^2$Laboratoire Matériaux et Phénomènes Quantiques, Sorbonne Paris Cité, Université Paris Diderot, CNRS UMR 7162, 75013 Paris, France}
\affiliation{$^3$Department of Physics, Technion Israel Institute of Technology, Haifa 32000, Israel}
\affiliation{$^4$Rafael Ltd., P.O. Box 2250, Haifa 32100, Israel}

\begin{abstract}

We investigate the topological properties of Fibonacci quasicrystals using cavity polaritons.
Composite structures made of the concatenation of two Fibonacci sequences allow investigating generalized edge states forming in the gaps of the fractal energy spectrum.
We employ these generalized edge states to determine the topological invariants of the quasicrystal. When varying a structural degree of freedom (phason) of the Fibonacci sequence, the edge states spectrally traverse the gaps, while their spatial symmetry switches: the periodicity of this spectral and spatial evolution yields direct measurements of the gap topological numbers. The topological invariants that we determine coincide with those assigned by the gap-labeling theorem, illustrating the direct connection between the fractal and topological properties of Fibonacci quasicrystals.

\end{abstract}

\pacs{03.65.Vf, 61.44.Br, 71.36.+c, 78.67.-n, 78.67.Pt}

\maketitle

Topology has long been recognized as a powerful tool both in mathematics and in physics. It allows identifying  families of structures which cannot be related by continuous deformations and are characterized by integer numbers called topological invariants.
A physical example where topological features  are particularly useful is provided by quantum anomalies, i.e. classical symmetries broken at the quantum level \cite{Jackiw72}, such as the chiral anomaly recently observed in condensed matter~\cite{Xiong15}.
From a general viewpoint, wave or quantum systems possessing a gapped energy spectrum, such as band insulators, superconductors, or 2D conductors in a magnetic field, can be assigned topological invariants generally called Chern numbers \cite{Chern79}. These numbers control a variety of physical phenomena: for instance in the integer quantum Hall effect, they determine the value of the Hall conductance as a function of magnetic field \cite{Klitzing80,Bellissard94}.
Such topological features related to Chern numbers have been explored in crystals \cite{Hasan10} and more recently in various artificial periodic lattices for cold atoms \cite{Goldman13,Atala13,Aidelsburger15}, acoustic waves \cite{xiao15} or photons \cite{Rechtsman13,Hafezi14,Hu15,Poli15,Skirlo15,Mittal16}.

Quasicrystals -- nonperiodic structures with long range order -- \cite{Shechtman84,Levine84} are another important class of systems exhibiting topological effects \cite{Simon82,Kunz86,Lang12,Kraus12,Kraus13,Ganeshan13}. In particular the topological edge states \cite{Zijlstra99,ElHassouani06,Pang10} of quasicrystals have been recently investigated in photonic systems \cite{Kraus12,Verbin13,Verbin15,Vignolo16} and exploited to implement topological pumping, a key concept of topology \cite{Kraus12}.
A paradigmatic example of quasicrystal is given by the 1D Fibonacci chain.
It presents a fractal energy spectrum which consists of an infinite number of gaps~\cite{Damanik08}.
A rather surprising and fascinating property is that each of theses gaps can also be assigned a topological number analogous to the aforementioned Chern numbers \footnote{Note that strictly speaking Chern numbers are invariants describing the topology of smooth Riemannian manifolds. In contrast, the Fibonacci chain and quasicrystals in general, cannot be ascribed such a smooth manifold but rather discrete graphs \cite{Anderson98} whose topological numbers are termed differently (K-theory invariants). For simplicity we shall pursue using indifferently the denomination Chern numbers or topological numbers.}: this constitutes the so-called gap-labeling theorem~\cite{Bellissard92}.
These integers can take $N$ distinct values, $N$ being the number of letters in the chain \cite{Dareau16}. 
Despite important advances on the topological properties of quasicrystals \cite{Simon82,Kunz86,Lang12,Kraus12,Kraus13,Ganeshan13,Kraus12,Verbin13,Verbin15,Vignolo16} the topological invariants have not yet been directly measured as winding numbers.

\begin{figure}[h]
\centering
\includegraphics[width=\columnwidth]{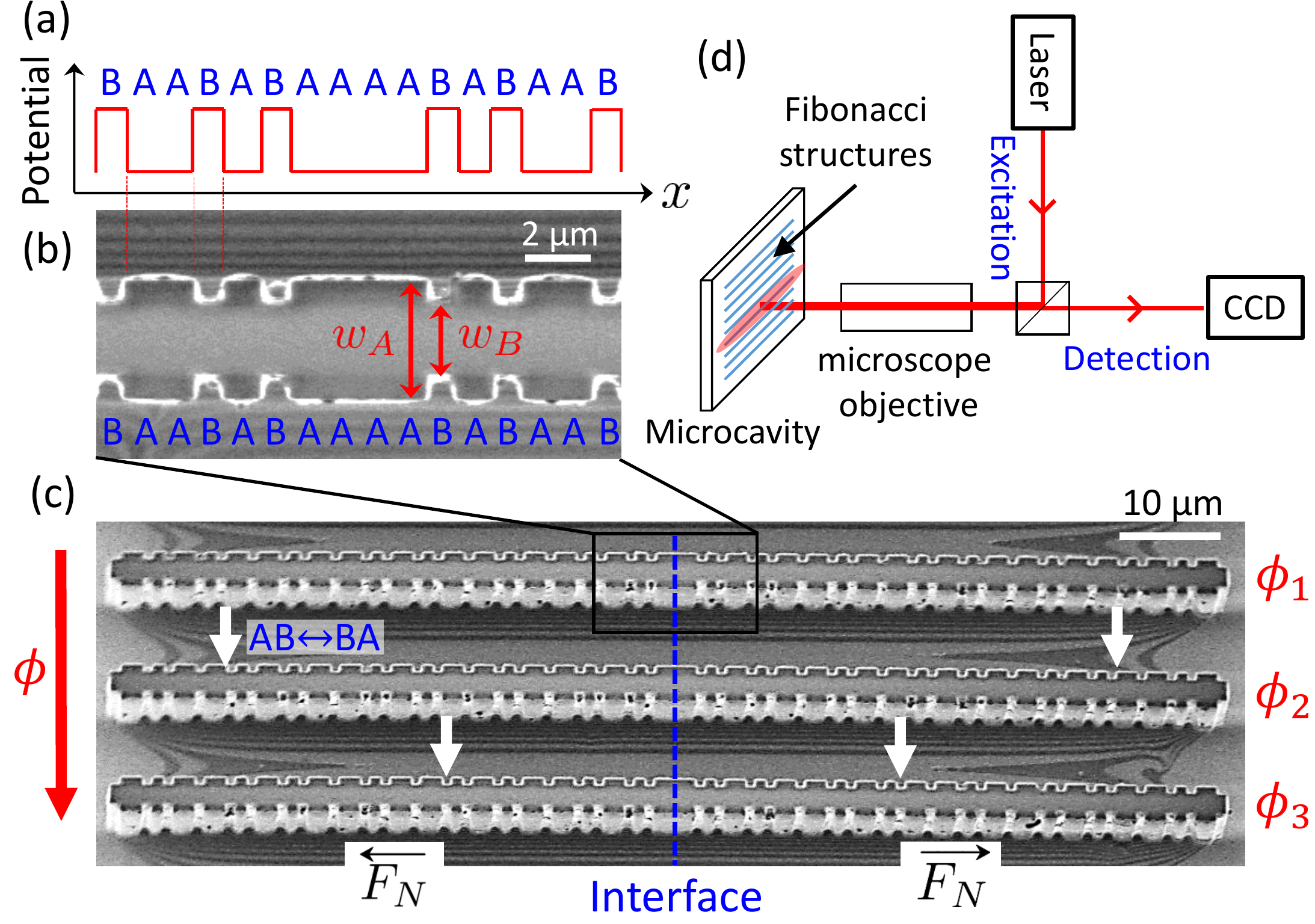}
\caption{
(a) Nominal potential energy corresponding to a laterally  modulated 1D cavity.
(b) SEM image of a portion of a 1D cavity reproducing the Fibonacci sequence (top view). The letters A and B correspond to two different widths of the cavity.
(c) SEM image showing the full view of 3 fabricated Fibonacci structures, corresponding to 3 different values of the phason $\phi$. Each structure consists of the concatenation $\Obif \Fibo$ of a Fibonacci sequence $\Fibo$ and its mirror symmetric $\Obif$.
Vertical white arrows indicate the position of local changes in the sequence introduced when scanning $\phi$.
(d) Schematics of the experimental setup.
}
\label{Fig1}
\end{figure}

The physical origin of topological numbers in a Fibonacci quasicrystal can be related to its structural properties~\cite{Levy15}. To understand this, let us introduce a general method to generate a Fibonacci sequence: it is based on the characteristic function
\begin{equation}
\chi_j = sign \, \left[ \cos \left( 2 \pi \, j \, \sigma^{-1} + \phi \right) - \, \cos \left( \pi \, \sigma^{-1} \right) \right],
\label{chi}
\end{equation}
proposed in \cite{Kraus12b},
which takes two possible values $\pm 1$, respectively identified with two letters A and B representing e.g. two different values of a potential  energy.
A Fibonacci sequence of size $N$ is a word $\Fibo (\phi)  \equiv \left[ \chi_1 \, \!\chi_2 \,\! \cdots \! \chi_N \right]$ formed by A and B letters. In Eq. \eqref{chi}, $\sigma = (1 + \sqrt{5} )/2$ is the golden mean and $\phi$ is a structural degree of freedom called phason, that can be continuously varied between $0$ and $2 \pi$.
The role of the phason has been experimentally investigated  in the frame of the Harper tight-binding model \cite{Kraus12,Kraus12b,Verbin13}.
In the case of the Fibonacci model $\phi$ allows selecting distinct finite segments along the infinite chain $\vv{F}_{\infty}$.
Sweeping $\phi$ over a $2 \pi$--period induces a series of $N$ independent local structural changes in the Fibonacci sequence $\Fibo$. Each change corresponds to the exchange of two letters ($AB\!\leftrightarrow\!BA$) at a given location of the sequence (see vertical arrows in Fig. \ref{Fig1}c).
Importantly, for two particular values of $\phi$ within a period, the Fibonacci sequence $\Fibo (\phi)$ becomes palindromic, i.e. it coincides with its mirror symmetric $\Obif (\phi)  = \left[ \chi_N \chi_{N-1} \cdots \! \chi_1 \right]$. In between these two values, $\phi$ drives a $\pi$--periodic symmetry cycle along the 1D structure.
When concatenating $\Fibo (\phi)$ with $\Obif (\phi)$, generalized edge states appear at the interface of the two mirror sequences, with properties tightly linked to $\phi$. Since topological invariants can always be written as winding numbers, it was predicted \cite{Levy15} that the Fibonacci topological numbers are measurable by counting how many times edge states traverse the gap while scanning the phason degree of freedom.

In this article we present a direct measurement of the topological invariants of a Fibonacci quasicrystal as spectral winding numbers. As an important consequence, we also relate these invariants to the observed change of spatial symmetry of generalized edge states, a robust feature largely insensitive to intrinsic disorder and other imperfections.
We employ cavity polaritons, quasiparticles  arising from the strong coupling between excitons confined in quantum wells and photons confined in a semiconductor microcavity \cite{Weisbuch92}. This photonic system allows
emulating a variety of Hamiltonians \cite{Bayer99,Kim11,Carusotto13,Baboux16} and characterizing the associated eigenstates both in the spectral and spatial domain \cite{Carusotto13}.
Here, we harness these features to emulate generalized edge states
forming in the gaps of the fractal Fibonacci spectrum.

\begin{figure*}[t]
\centering
\includegraphics[width=\textwidth]{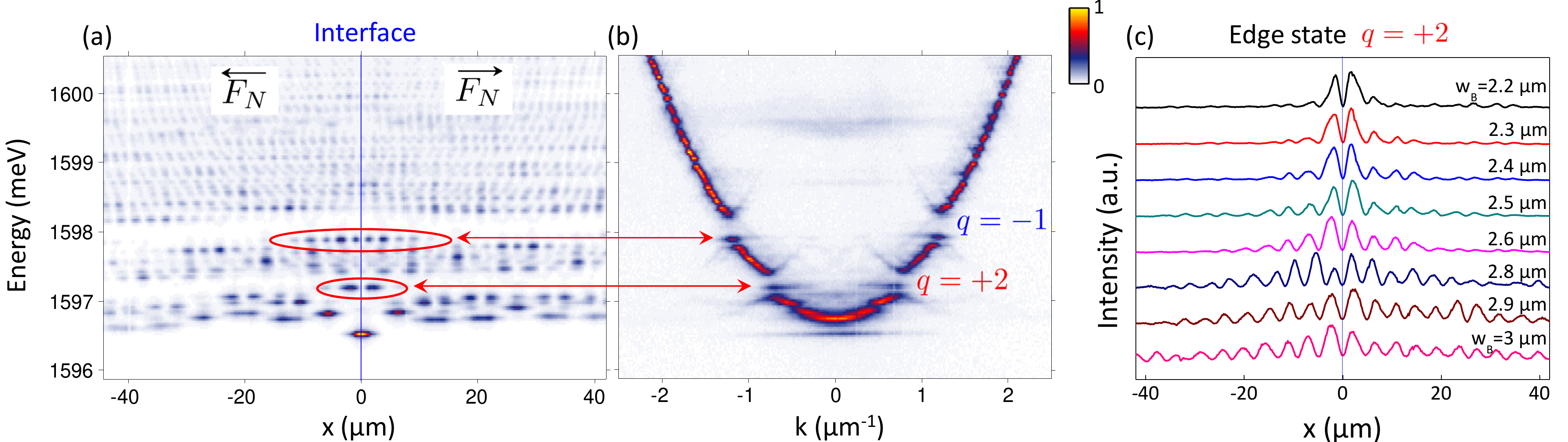}
\caption{(a)-(b) Energy-resolved emission of a Fibonacci structure in real space (a) and  in momentum space (b), for a given value of the phason ($\phi=0.62 \pi$). Edge states are visible in the two lowest main energy gaps, characterized by $q=+2$ and $q=-1$. These states are localized at the interface ($x=0$) between the $\Obif$ and $\Fibo$ Fibonacci sequences.
(c) Spatial profile of the edge state of gap $q=+2$  measured in a series of structures of same A-letter width $w_A=3.5$ $\mu$m but various B-letter width $w_B$, yielding different contrasts for the Fibonacci potential.
}
\label{Fig2}
\end{figure*}

To explore the edge states, we design concatenated structures $\Obif \Fibo$ made of the juxtaposition  of a given Fibonacci sequence and its mirror symmetric (Fig. \ref{Fig1}c). The interface defines a Fabry-Perot cavity of zero geometric length but finite round trip phase $\theta_{\rm cav}$ due to the reflexion between $\Fibo$ and $\Obif$ (see Supplemental Material \cite{Note3}).
Thus the edge states will appear at energies $E_{\rm gap}(\phi)$, implicitly  determined by the resonance condition 
\begin{equation}
\theta_{\rm cav} (E_{\rm gap},\phi)= 2\pi m \, ,
\label{eq2}
\end{equation}
with $m \in  \mathbb{Z}$. As detailed in \cite{Levy15}, $\theta_{\rm cav}$ is periodic in $\phi$ with a period $\frac{\pi}{|q|}$, where $q$ is an integer, the topological invariant of the considered gap. Therefore, the evolution of the energy of the edge states when increasing $\phi$ reflects directly the topological properties of the gap in which they appear.

To fabricate these structures, we process a planar microcavity (of nominal $Q$ factor $70000$) grown by molecular beam epitaxy. The cavity consists in a $\lambda / 2$ Ga$_{0.05}$Al$_{0.95}$As layer surrounded by two Ga$_{0.8}$Al$_{0.2}$As/Ga$_{0.05}$Al$_{0.95}$As Bragg mirrors with 28 and 40 pairs in the top and bottom mirrors respectively. Twelve GaAs quantum wells of width 7 nm are inserted in the structure, resulting in a $15$ meV Rabi splitting.
Quasi-1D cavities (wires) are realized using electron beam lithography and dry etching. The lateral width of these wires is modulated quasi-periodically, as shown in the Scanning Electron Microscopy (SEM) image of Fig. \ref{Fig1}b. The modulation consists in two wire sections A and B of same length $a=1\,\mu$m, but different widths $w_A$ and $w_B$. The width modulation induces an effective 1D potential for the longitudinal motion of polaritons (Fig. \ref{Fig1}a), that follows the desired Fibonacci sequence. We chose $N = 55$ letters for the Fibonacci sequences and thus $110$ letters for the concatenated structures.
We have fabricated on a single sample the $N=55$ concatenated structures corresponding to all possible values of $\phi$ producing a structural change in the sequence. Figure \ref{Fig1}c shows a subset of three fabricated structures; the position of the interface between the mirror sequences $\Obif$ and $\Fibo$ is indicated by a vertical line. The exciton-photon detuning is of the order of $-20$ meV for all experiments.

To study the polariton modes in these quasiperiodic structures, we perform low temperature (10 K) micro-photoluminescence experiments (see Fig. 1d).  Single structures are excited non-resonantly at low power, using a CW monomode laser at 740 nm. The excitation spot covers a $80\,\mu m$-long region centered on the interface.
The emission is collected with a $0.5$ numerical aperture objective and focused on the entrance slit of a spectrometer coupled to a CCD camera. Imaging the sample surface or the Fourier plane of the collection objective allows studying the polariton modes either in real or momentum space.

Figure 2 shows the photoluminescence spectrum in real space [Fig. \ref{Fig2}(a)] and in momentum space [Fig. \ref{Fig2}(b)] of a Fibonacci structure with $\phi=0.6 \, \pi$, $w_A=3.5\,\mu$m and $w_B=2.2\,\mu$m.
The energy spectrum shows an alternation of minibands and gaps.
The emission in real space allows identifying two types of modes in the spectrum: modes delocalized over the whole structure, and modes localized at the interface between the $\Obif$ and $\Fibo$ sequences ($x=0$). The delocalized modes form minibands, as can be seen in the momentum space emission: these modes are bulk modes forming a fractal energy spectrum characteristic of the Fibonacci quasicrystal \cite{Tanese14}.
We can identify the main gaps of the spectrum by applying the gap-labeling theorem \cite{Bellissard92}, which predicts $k=\frac{\pi}{a} (p+q \sigma^{-1})$ for the wavevector position of the gaps~\footnote{This property, exact in the limit of infinite chains, constitutes an excellent approximation for the considered structure length, as shown in Ref. \cite{Tanese14}.}. Here, $p$ and $q$ are integers, with $q$ being the gap topological number \cite{Bellissard92}.
From the momentum space spectrum of Fig. \ref{Fig2}b we extract
$[p,q]=[-1,2]$ for the lower main energy gap, and $[p,q]=[1,-1]$ for the higher main energy gap.

In addition to the bulk states, we observe states that are localized at the interface between $\Obif$ and $\Fibo$. One of them (at energy $\sim 1596.5$ meV) lies below the bulk band structure and is thus topologically trivial \footnote{See Supplemental Material for details on the Fabry-Perot interpretation and additional data}. Two other localized states (encircled) lie within the widest spectral gaps: these are the expected topological edge states.
Their spatial localization around the interface depends on the contrast of the Fibonacci quasiperiodic potential. Figure \ref{Fig2}c shows the spatial profile (squared modulus of the wavefunction, $|\psi(x)|^2$) of the edge state of the gap $q=+2$, measured for a series of structures of same A-letter width  $w_A=3.5$ $\mu$m, but various B-letter widths $w_B$. As $w_B$ decreases, the potential contrast (amplitude of the steps in Fig. \ref{Fig1}a) increases: this leads to wider spectral gaps, and thus to a stronger spatial localization of the wavefunction.

\begin{figure}[h]
\centering
\includegraphics[width=1\columnwidth]{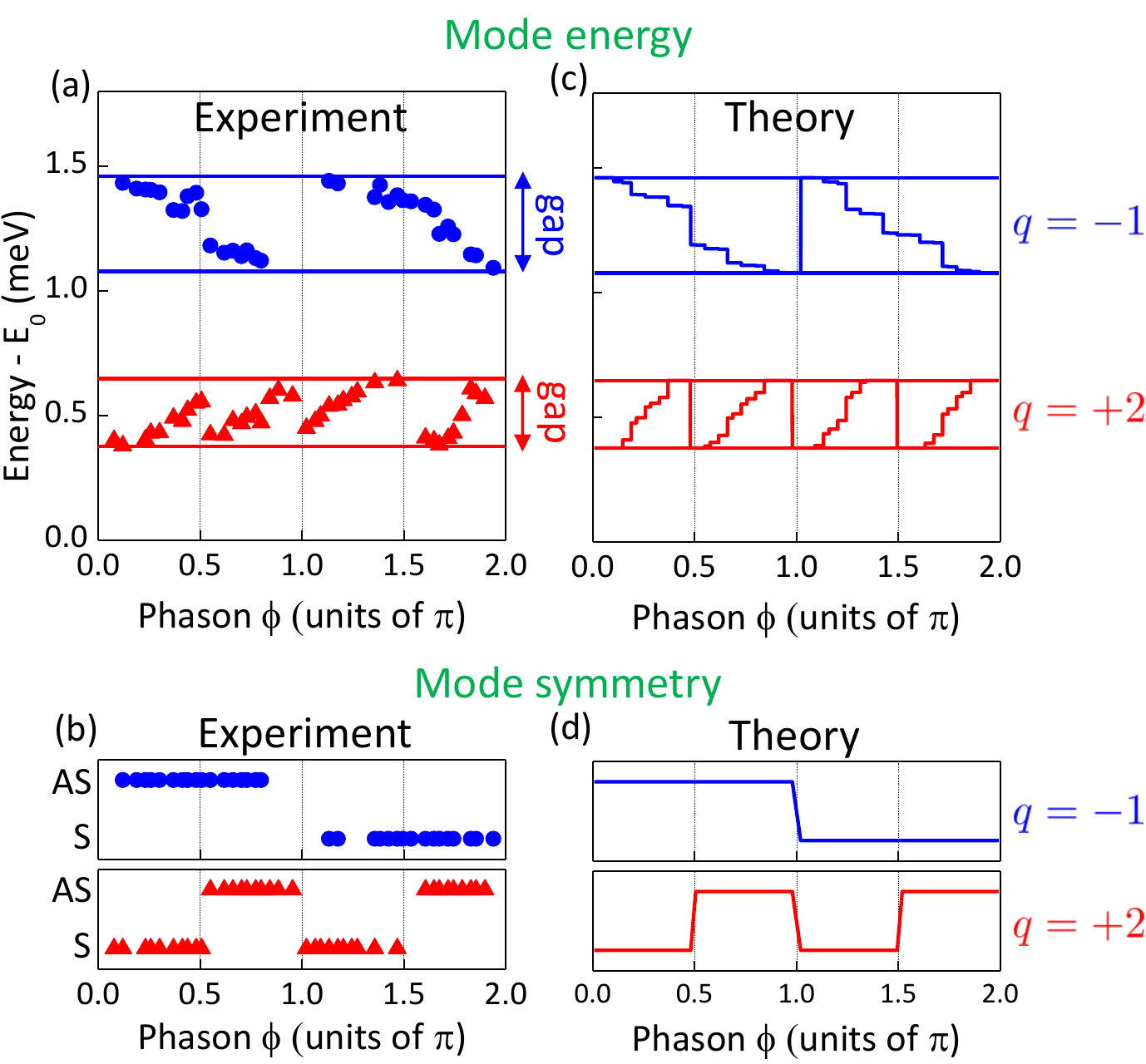}
\caption{
(a) Measured energy of the edge states of gaps $q=+2$ et $q=-1$ as a function of the phason $\phi$. $E_0$ denotes the energy of the lowest bulk mode, and the solid lines indicate the gap boundaries.
(b) Corresponding spatial symmetry of the edge states. When scanning $\phi$, the wavefunctions evolve from symmetric (S) to anti-symmetric (AS) with respect to the interface ($x=0$).
(c) Relative spectral position within the gaps of the two considered edge states, obtained from scattering matrix calculations. (d) Calculated symmetry of the two considered edge states.
}
\label{Fig3}
\end{figure}

To explore the topological properties of the Fibonacci sequences by means of these edge states, we will now monitor their evolution when varying the phason degree of freedom $\phi$. We investigate a full set of $N=55$ structures with $w_A=3.5\,\mu$m and $w_B=2.4\,\mu$m.
For each structure, we perform spectroscopic measurements similar to Fig. \ref{Fig2}a-b, and extract the energy of the two edge states with respect to the lowest bulk energy mode (bottom of the parabola, energy $E_0$). The results are plotted in Fig. \ref{Fig3}a, where the gap boundaries are indicated by the horizontal lines. Numerical calculations based on a scattering matrix approach \cite{Levy15} are presented for comparison in Fig. \ref{Fig3}c. Note that when $\phi=0$ or $\pi$, no edge state is observed. Indeed, for these particular values of $\phi$, the $\Fibo$ sequence is palindromic and  $\Obif \,\Fibo$ effectively reduces to a single Fibonacci sequence of size $2N$. Hence there is no interface cavity and thus no edge state.

As clearly seen in Fig. \ref{Fig3}a, while scanning the phason $\phi$ the states perform piecewise spectral traverses inside the gaps.
The number and the direction of the traverses yield a direct determination of the winding number \cite{Levy15}:
\begin{align}
\mathcal{W}
=\frac{1}{2 \pi} \int_0^{2\pi} {\frac{d \theta_{\rm cav}}{d \phi}} d \phi \,
 = \frac{1}{2 \pi} \int_0^{2\pi} {\frac{d \widetilde{\delta}}{d \phi}} d \phi = 2q
\end{align}
where $\widetilde{\delta} \equiv \frac{ E_{\rm gap}-E_-}{\Delta_g}$ is the spectral position of the edge state within the gap, with $\Delta_g$ being the gap width and $E_-$ the energy of the gap lower boundary \footnote{Since the quasicrystal is an effectively discrete manifold, the phase $\theta_{\rm cav}$ defined in Eq. \eqref{eq2} can take only a finite set of discrete values equal to the Fibonacci number $F_N$. Thus, strictly speaking the winding number $\mathcal{W}$ should be written as a discrete sum. As a convenient notation which results from the use of a continuous variable $\phi$ in Eq. \eqref{chi}, we replace this discrete sum by an integral.}.

The direction and periodicity of the observed traverse is different for the two edge states we consider: the lower energy state traverses $4$ times upwards (winding number $\mathcal{W}=+4$), while the higher energy state traverses $2$ times downwards ($\mathcal{W}=-2$), when $\phi$ spans a full period $[0,2\pi]$.
This winding $\mathcal{W}=2q $ of the edge states allows for a direct determination of the gap topological numbers.
We deduce $q=+2$ for the lower energy state and $q=-1$ for the higher energy state. These values obtained from the winding of the \textit{edge} states are fully consistent with those previously determined from the \textit{bulk} band structure (gap-labeling theorem), illustrating the existence of a bulk-edge correspondence in the quasiperiodic system.

\begin{figure}[h]
\centering
\includegraphics[width=0.95\columnwidth]{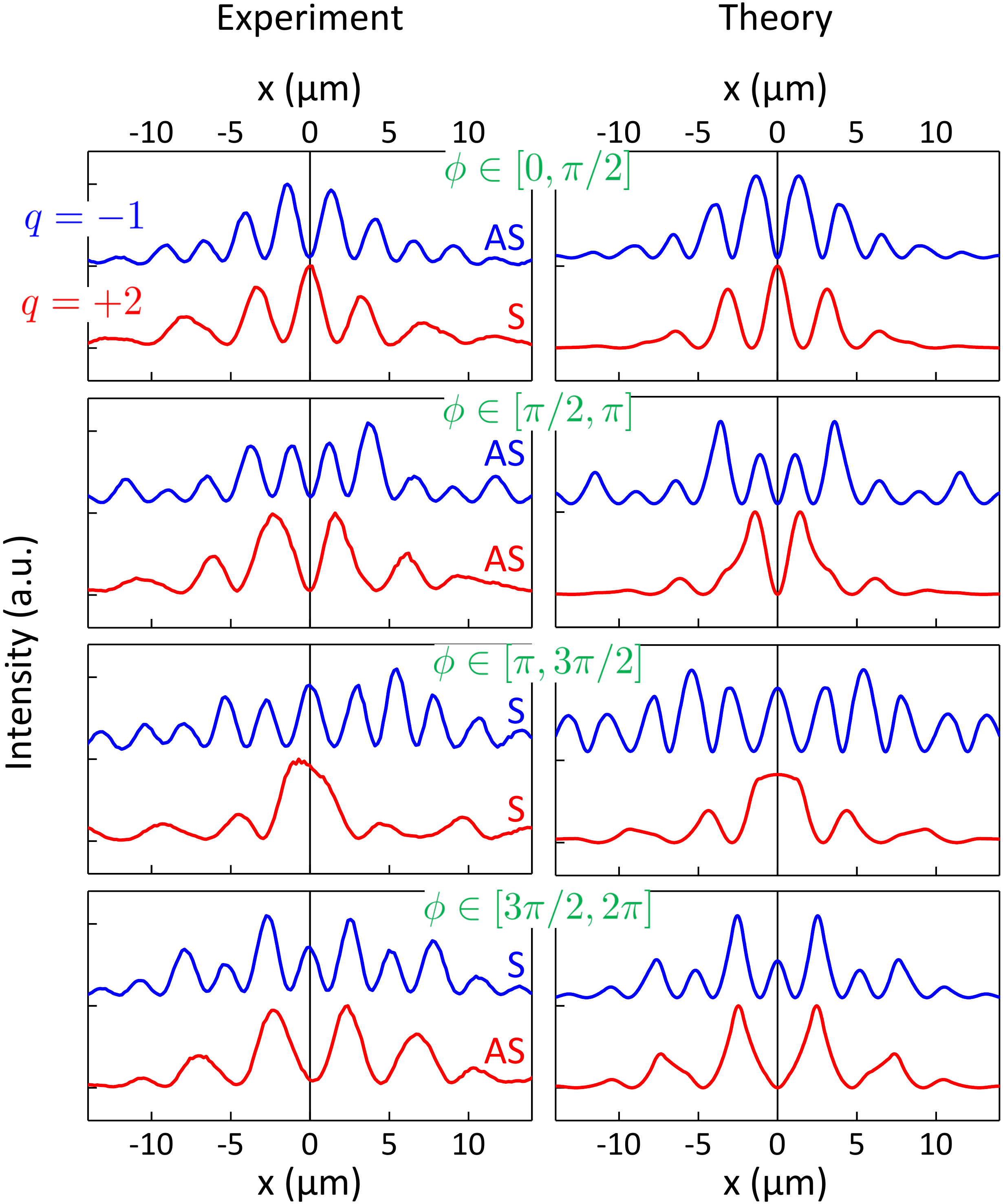}
\caption{
Measured (left) and calculated (right) spatial profile of the $q=+2$ and  $q=-1$ edge states for values of $\phi$ taken in 4 different quadrants: $[0,\pi/2]$, $[\pi/2, \pi]$, $[\pi, 3\pi/2]$ and $[3\pi/2, 2 \pi]$. The mode spatial structure switches from symmetric (S) to antisymmetric (AS) with respect to the interface ($x=0$), with a direction and periodicity yielding a direct measurement of the topological invariants of the quasicrystal.
The blue lines are shifted vertically to improve their visibility.}
\label{Fig4}
\end{figure}

We now show that the topological invariants of the quasicrystal are not only measurable as winding numbers of the edge states, but they can also be directly retrieved from the spatial symmetry of the corresponding wavefunctions.
Figure \ref{Fig4} (left column) shows the measured profile of the $q=+2$ and  $q=-1$ edge states for values of $\phi$ taken in the $4$ successive quadrants: $[0,\pi/2]$, $[\pi/2, \pi]$, $[\pi, 3\pi/2]$ and $[3\pi/2, 2 \pi]$.
The states either show a maximum or a minimum intensity at the interface ($x=0$), corresponding to either a node or an antinode of the wave function. In the frame of a Fabry-Perot model \cite{Levy17} we will denote these two cases as symmetric (S) and anti-symmetric (AS) respectively. Because of experimental imperfections the states are not perfectly S or AS but we shall employ this convenient terminology. We observe that the $q=+2$ state (red) switches symmetry in each quadrant, while the $q=-1$ state (blue) keeps the same symmetry in the first two quadrants before switching to the opposite symmetry in the last two quadrants.
The symmetry index (S or AS) is reported in Fig. \ref{Fig3}b for all values of $\phi$, and compared to theory in Fig. \ref{Fig3}d.
Comparing this behavior to the spectral features reported in Fig. \ref{Fig3}a-c, we observe that symmetry flips are exactly synchronized with the spectral traverse of the states: they occur in between two successive traverses.
Their periodicity thus allows determining the absolute value of the topological numbers.
In addition, we observe that the sign of $q$ is reflected in the direction of the symmetry flips within a period: the wavefunction of the $q=-1$ state  switches from AS to S while that of the  $q=+2$ state switches from S to AS.
These experimental observations can be fully accounted for by a Fabry-Perot interpretation. The resonance condition of Eq. \eqref{eq2}, besides giving an accurate prediction for the gap states energies as shown in Figs. \ref{Fig3}a and \ref{Fig3}c, also encodes information on the mode symmetry. Indeed, for a fixed cavity length the occurrence of a node or antinode at the center is fully determined by the parity of the integer $m$. The spatial symmetry of the topological states thus switches at each spectral traverse \cite{Note3}. This can be seen in the right column of Fig. \ref{Fig4}, showing the calculated spatial profile of the two gap states under study, in all four quadrants.
These features demonstrate that, in contrast to previous studies using normal edges (interface with vacuum) \cite{Kraus12,Verbin15}, the use of generalized edges yields an additional degree of freedom (the symmetry index) that can be used to directly measure the topological invariants of the quasicrystal. This novel method is independent from the one based on spectral windings. It should thus prove useful in other physical platforms where the spectral degrees of freedom are more difficult to access.

In summary, we have investigated the topological properties of 1D Fibonacci polaritonic quasicrystals.
We image generalized edge states forming in the gaps of the fractal energy spectrum. The behavior of these edge states upon varying a structural degree of freedom (phason) allow a direct determination of the topological invariants of the quasicrystal.  The latter coincide with the bulk invariants assigned by the gap-labeling theorem, illustrating the direct connection between the fractal and topological properties of Fibonacci quasicrystals.
This work demonstrates a novel approach to determine  topological features of quasicrystals by means of a direct measurement of winding numbers and of the corresponding symmetry flips of the wavefunctions. Taking advantage of the matter part of polaritons, it could be extended to probe the interplay of topology and interactions in quasicrystals \cite{Freedman06} and potentially realize strongly correlated topological phases \cite{Wen95}. Furthermore, the building of cavities with topological mirrors could allow the investigation of the topological Casimir effect \cite{Bordag09} in a well-controlled platform.

\bigskip

\textbf{Acknowledgments}: This work was supported by the Israel Science Foundation Grant No. 924/09, by the Agence Nationale de la Recherche projects \textit{Quandyde} (Grant No. ANR-11-BS10-001) and \textit{Quantum Fluids of Light} (Grant No. ANR-16-CE30-0021), the French \textit{RENATECH} network, the European Research Council grant \textit{Honeypol} and the EU-FET Proactive grant \textit{AQuS} (Project No. 640800).

\bibliography{Biblio}


\clearpage
\onecolumngrid
\begin{center}
\textbf{\large Supplemental Material --- Measuring topological invariants \\
from generalized edge states in polaritonic quasicrystals}
\end{center}
\twocolumngrid

\setcounter{equation}{0}
\setcounter{figure}{0}

\renewcommand{\theequation}{S\arabic{equation}}
\renewcommand{\thefigure}{S\arabic{figure}}

\newcommand{\be}{\begin{equation}}
\newcommand{\ee}{\end{equation}}
\newcommand{\bea}{\begin{eqnarray}}
\newcommand{\eea}{\end{eqnarray}}

\section{Fabry-Perot model of the topological edge states}

\begin{figure}[h]
\includegraphics[width=1\columnwidth]{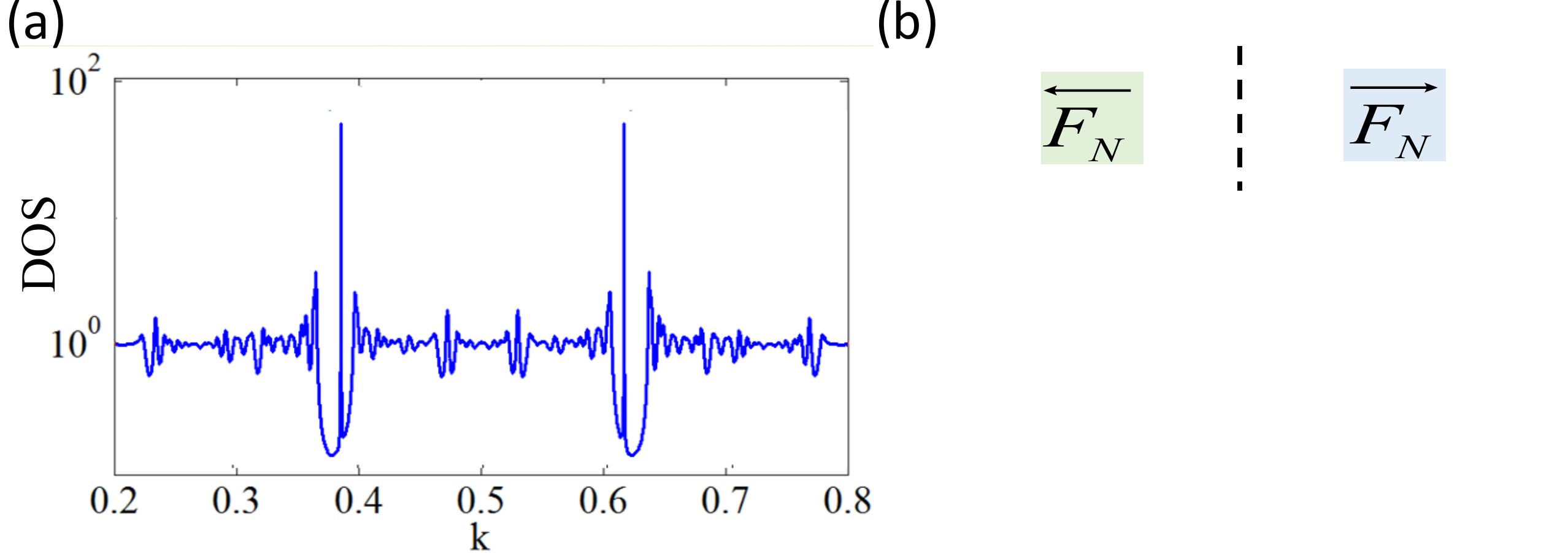}
\caption{Appearance of gap states in the Fibonacci based structure $\Obif \Fibo$ with $N=55$. (a)  Density of states spectrum DOS$(k)$ with interface states appearing in the gaps (here at normalized wavevectors $k=0.39$ and $k=0.61$). (b) Spatial arrangement of the structure $\Obif \Fibo$ indicated by yellow and magenta bars (corresponding to letters A and B), with a representation of the spatial profile of the gap state at $k=0.39$.
\label{Fig5}}
\end{figure}

The effective Fabry-Perot model derived here is for a true 1D system, and is given in the language of the scattering theory, as originally proposed in Ref. [35]. The basic idea is to regard the interface between the Fibonacci sequences $\Obif(\phi)$ and $\Fibo(\phi)$ as a virtual Fabry-Perot cavity, and correspondingly consider the chains $\Obif(\phi)$ and $\Fibo(\phi)$ as mirrors. In this perspective, the generalized edge states localized at the interface between the two Fibonacci chains are interpreted as Fabry-Perot resonances (see Fig. \ref{Fig5}).

Qualitatively speaking, the Fibonacci chains $\Obif(\phi)$ or $\Fibo(\phi)$ which have a  gapped spectrum are mirrors for specific frequencies, due to the high reflectance values at the spectral gaps. These mirrors are not standard, since they provide a frequency-dependent phase shift upon reflection due to multiple reflections. This phase shift, as we will show now, allows to treat an interface between two such structures as a virtual (not geometric) cavity length.

The standard Fabry-Perot resonance condition for a cavity of length $L$, given by  
\be
2L/\lambda_{m}= m,  \, \, \, m\!\in\!\mathbb{Z},
\label{FP_res}
\ee
($\{\lambda_{m}\}$ being a discrete set of resonant wavelengths), is a basic constructive interference condition. Therefore, it may always be written in terms of the winding of a phase: the cavity phase $\theta_{\rm cav}$, representing the total round-trip phase inside the cavity and defined by
\be
\theta_{\rm cav}\left(k,L\right)\equiv\frac{4\pi L}{\lambda\left(k\right)},
\label{FP_res2}
\ee
where $\lambda(k)\!=\!2\pi/k$ is the wavelength. This gives a resonance condition equivalent to Eq. \eqref{FP_res}, namely
\be
\theta_{\rm cav}(k_{m})=2\pi m\;,\;m\!\in\!\mathbb{Z}.
\ee 

Here we follow the very same argument, but in the reverse order. We begin with the structure $\Obif \Fibo$, which is a cavity of zero (geometrical) length. The cavity phase is however non-zero, as $\Obif$ and $\Fibo$ both yield a non-zero phase upon reflection. If we define the reflected phase shift  for the left boundary of $\Fibo$ as $\thetaa_{\rm left}\left(k\right)$, then the cavity phase for $\Obif\Fibo$ is $2\thetaa_{\rm left}\left(k\right)$. Now, similarly to Eq. \eqref{FP_res2}, a virtual cavity length is defined as
\be
{\cal L}\left(k,\theta_{\rm cav} \right) \equiv \frac{\lambda(k)}{4\pi}\,\theta_{\rm cav},
\label{effect_length}
\ee
with resonant interface states occurring at gap frequencies satisfying the Fabry-Perot condition
\be
2{\cal L}(k_{m})/\lambda(k_{m})\!=\! m, \, \, \, m\!\in\!\mathbb{Z}.
\label{FP_res3}
\ee
The model then predicts that for every value of $\lambda(k_{m})$ lying within a spectral gap, a new interface state will exist. 

This reasoning is utilized in Eqs. (2) and (3) of the article directly through $\theta_{\rm cav}$, to define the topological winding number of the generalized edge state when the phason $\phi$ is scanned. In that case, the scanning of $\phi$ monotonically drives the cavity phase (at gap frequencies) such that the resonant states monotonically change frequency with $\phi$. This reasoning fully supports the experimental observation of the topological winding of generalized edge states as a function of $\phi$ in  momentum space, as evidenced in Fig. 3a,b of the letter.
We now describe a Fabry-Perot property in real space that helps to further clarify the interplay between the topological and the Fabry-Perot properties of the generalized edge states.

\begin{figure}
\includegraphics[width=0.8\columnwidth]{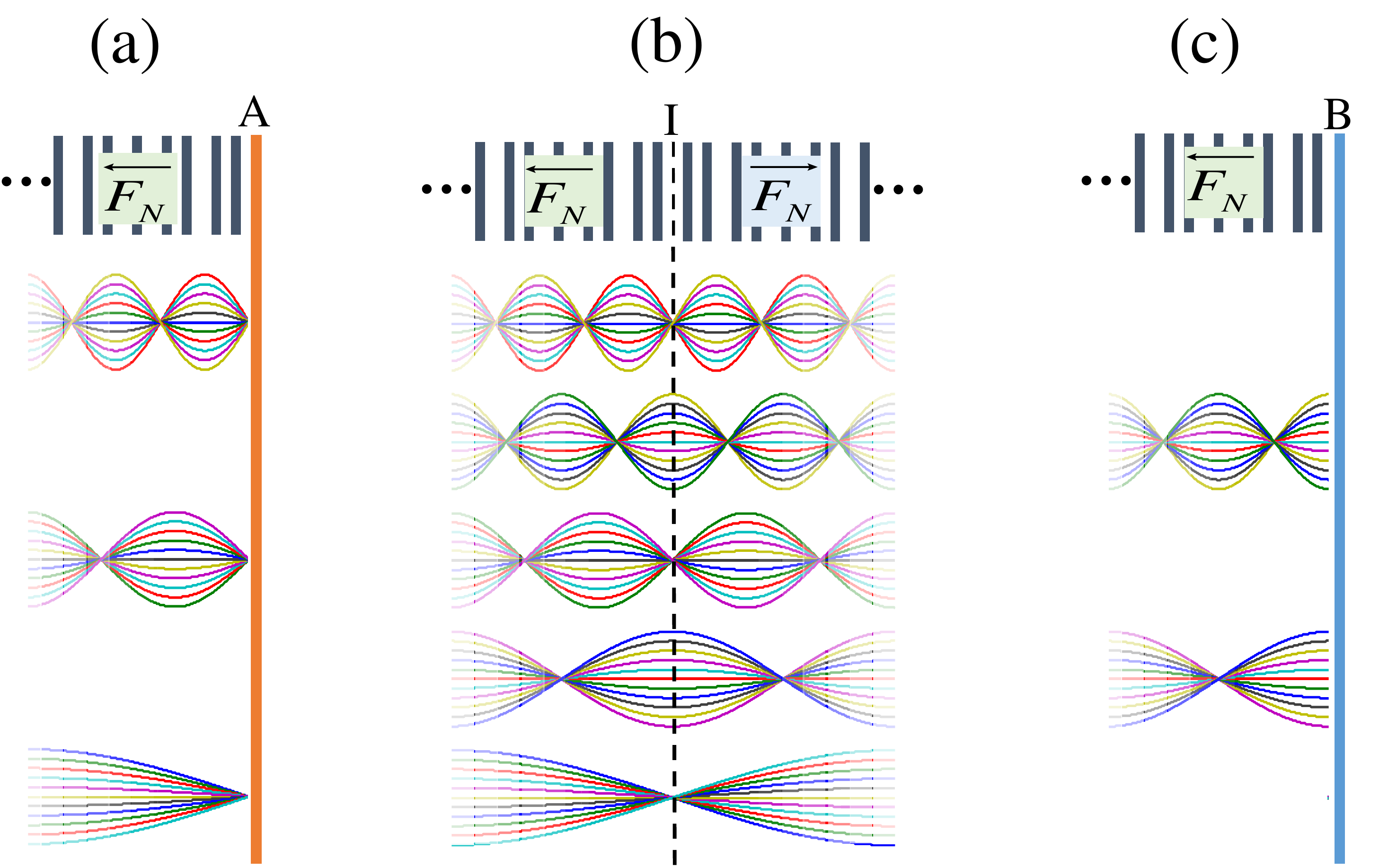}
\caption{Different boundary conditions for the Fibonacci chain.
(a) and (c): The Fibonacci chain $\Obif$ with $N=55$ bounded from the right (a) by a metallic mirror  (orange line), or (c) by a continuum with refractive index smaller than that of the chain (blue line). The presence of a node or antinode of the gap states at the boundary is here rigidly imposed by the boundary conditions.
(b) The unfolded structure $\Obif \Fibo$, based on the same structure, hosts generalized edge states which can possess either a node or an antinode at the interface (dotted line).}
\label{Fig6}
\end{figure}

\begin{figure}
\includegraphics[width=0.9\columnwidth]{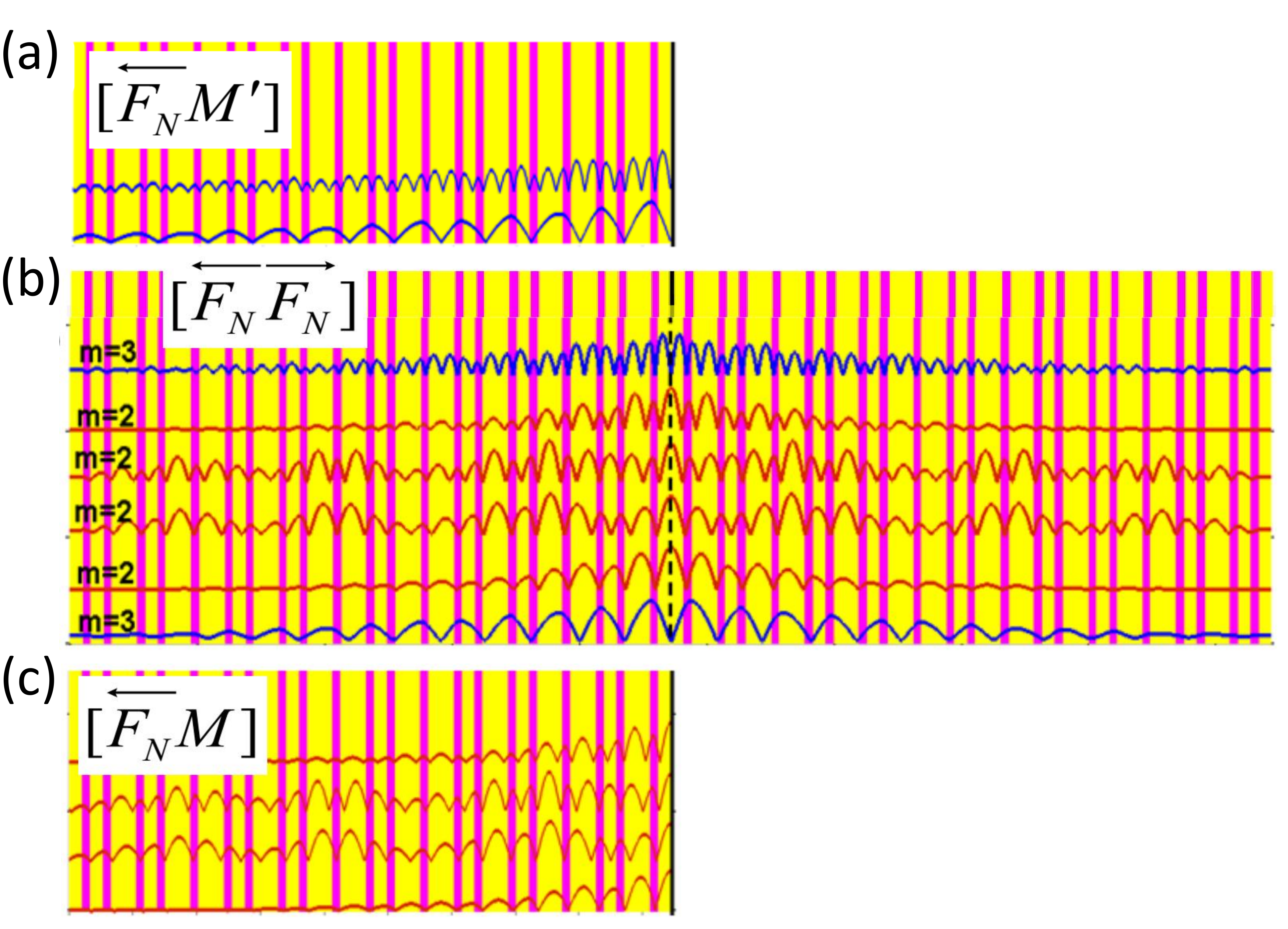}
\caption{Real space profile of selected gap states for the structures depicted in Fig. \ref{Fig6}: (a) The $\Obif$ chain with metallic reflective boundary, (b) the unfolded structure $\Obif \Fibo$ and (c) the $\Obif$ chain with a refractive-index mismatch boundary.
\label{Fig8}}
\end{figure}

The structure $\Obif\Fibo$ is described in the article as a host for generalized edge states, in contrast to previous studies considering the interface with the vacuum. To clarify this, we consider the $\Obif$ chain bounded from the right by a perfect mirror. Waves traveling through the structure and towards the mirror plane are reflected back into the chain, experiencing the quasiperiodic modulation in the reverse order ($\Fibo$). An equivalent version of this setup consists in removing the mirror and unfolding the chain with respect to the mirror plane, resulting in the structure considered in the article. The equivalence between these two boundary conditions is however not total. In particular, the structure $\Obif\Fibo$ may host interface states which are both spatially symmetric and anti-symmetric with respect to the interface, namely with a node or an anti-node at the interface, as seen in Fig. \ref{Fig6}b.
In contrast to that, a single $\Obif$ chain bounded by a perfect mirror can host only one type of states: with a node on the mirror in the case of a metallic mirror, as shown in Fig. \ref{Fig6}a, or with an antinode in the case of a mismatched mirror (continuum with refractive index smaller than that of the chain), as shown in Fig. \ref{Fig6}c.

The fact that the generalized-edge scheme allows all possible interface states yields an additional degree of freedom to probe the topological content of the states. Indeed, topological information may be extracted from the real space properties of the edge states through the parity of the Fabry-Perot integer, $m$ in Eq. \eqref{FP_res3}. In the usual Fabry-Perot picture for phase-conserving perfect mirrors, odd and even values of $m$ alternate between anti-symmetric and symmetric states with respect to the mid-cavity coordinate (with a node and an antinode at mid-cavity) respectively. This result is also true for our Fabry-Perot cavity with a $\phi$ dependent virtual length. This leads to two possible predictions. 

Firstly, for a  given value of $\phi$, edge states residing in the various gaps have different spatial symmetries completely predictable by the Fabry-Perot model through the parity of $m$. 
For instance, Fig. \ref{Fig8} shows that modes with $m=2$ (red) are symmetric with respect to the interface, while modes with $m=3$ (blue) are antisymmetric. Again, we here see that the general-edge scheme (\ref{Fig8}b) hosts the union of all edge states of the metallic and index-mismatched boundary conditions (\ref{Fig8}a,c) taken together.

Secondly, for a  given gap, generalized edge states traverse the gaps as a function of $\phi$ due to a monotonic change in the cavity phase. This means that when an edge state merges with one band-edge and a new state bifurcates from the other band-edge, then the parity of $m$ flips and so does the spatial symmetry of the edge state. This is the theoretical argument supporting the experimental observation reported in Fig. 4 of the article.
These features demonstrate that the use of generalized edges yields an additional degree of freedom (the symmetry index) that can be used to directly measure the topological invariants of the quasicrystal.

\section{Topologically trivial modes of the polaritonic quasicrystals}

In this section we show that our polaritonic structures also hosts topologically trivial modes, that can be clearly distinguished from the topological modes described above.
An example is given in Fig. 2a,b of the article. Among the states that are localized at the interface between the $\Obif$ and $\Fibo$ sequences, one of them (at energy $\sim 1596.5$ meV) lies below the bulk band structure and is thus topologically trivial.

\begin{figure}[h]
\centering
\includegraphics[width=\columnwidth]{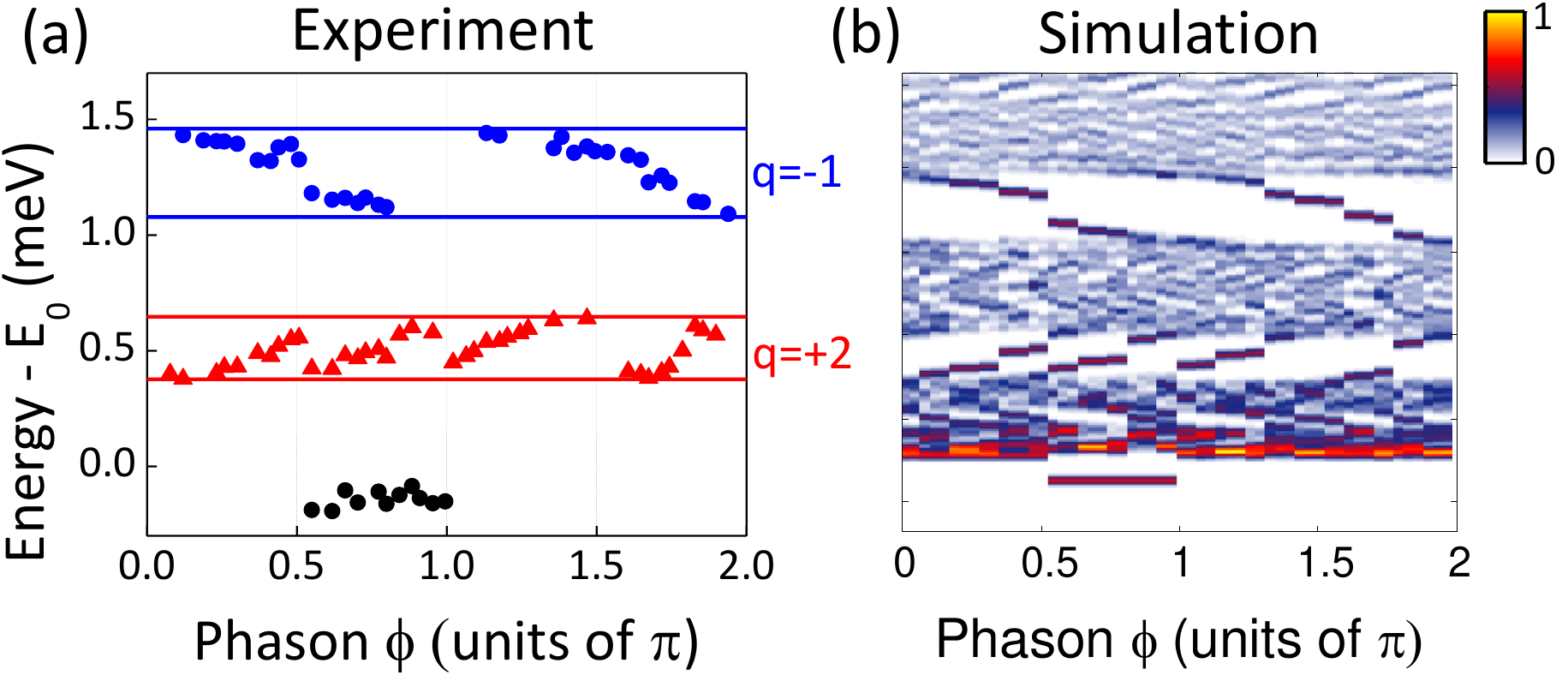}
\caption{(a) Measured energy of the interface states (for the structure considered in Figs. 2,3,4 of the main text) as a function of the phason $\phi$. The topologically trivial mode is indicated with black circles. $E_0$ denotes the energy of the lowest bulk mode, and the solid lines indicate the gap boundaries.
(b) Calculation of the whole energy spectrum from the 2D Schr\"{o}dinger equation, as a function of $\phi$.
}
\label{Fig9}
\end{figure}

This mode only appears for a particular quadrant of $\phi$ (between $\pi/2$ and $\pi$), for which the sequence of letters at the interface is AAAA. Since in our samples, the letter A corresponds to a lower potential value, this AAAA sequence forms a spatially extended potential well, and the mode that is seen is a bound mode of this well. To verify that this mode is topologically trivial, we can monitor its spectral evolution as a function of $\phi$, both in the experiment and the simulation (2D Schr\"{o}dinger calculation) as shown in Fig. \ref{Fig9}. We observe that the mode only appears in the quadrant $[\pi/2, \pi]$ and keeps a constant energy within the experimental error: it does not perform any spectral traverse as the phason is scanned, which confirms its topologically trivial nature.

\end{document}